\documentclass[aps,pra,twocolumn,superscriptaddress]{revtex4}

\pdfoutput=1

\usepackage{amsmath}
\usepackage{amssymb}
\usepackage{amsbsy}
\usepackage{makeidx}
\usepackage{epsf}
\usepackage{graphicx}
\usepackage{amsfonts}
\usepackage{subfigure}
\usepackage{epstopdf}

\begin{document}
\title{Persistence of equilibrium states in an oscillating double-well potential}
\author{H.\ Jiang}
\author{H.\ Susanto}
\affiliation{School of Mathematical Sciences, University of Nottingham, University Park, Nottingham, NG7 2RD, UK}
\author{T.M.\ Benson}
\affiliation{Electrical Systems and Optics Division, Faculty of Engineering, University of Nottingham, University Park, Nottingham, NG7 2RD, UK}
\author{K.A.\ Cliffe}
\affiliation{School of Mathematical Sciences, University of Nottingham, University Park, Nottingham, NG7 2RD, UK}


\begin{abstract}
We investigate numerically parametrically driven coupled nonlinear Schr\"odinger equations modelling the dynamics of coupled wavefields in a periodically oscillating double-well potential. The equations describe among other things two coupled periodically-curved optical waveguides with Kerr nonlinearity or horizontally shaken Bose-Einstein condensates in a double-well magnetic trap. In particular, we study the persistence of equilibrium states of the undriven system due to the presence of the parametric drive. Using numerical continuations of periodic orbits and calculating the corresponding Floquet multipliers, we find that the drive can (de)stabilize a continuation of an equilibrium state indicated by the change of the (in)stability of the orbit. Hence, we show that parametric drives can provide a powerful control to nonlinear (optical or matter wave) field tunneling. Analytical approximations based on an averaging method are presented. Using perturbation theory the influence of the drive on the symmetry breaking bifurcation point is discussed.
\end{abstract}

\maketitle

\section{Introduction}

Parametric drives, i.e.\ external drives that are periodic in time and depend on the system variables, have been used as a means to control and maintain a system out-of-equilibrium \cite{chen98,cros93}. The generation of standing waves when a liquid layer is subjected to vertical vibration, known as Faraday waves, is among the classical studies of parametrically driven instabilities \cite{fara31}. The vertical vibration can sustain spatially localized, temporally oscillating structures, commonly referred to as oscillons, such as  in granular materials \cite{clem96,umba96}, Newtonian \cite{liou96,arbe00} and non-Newtonian fluids \cite{liou99}. Relatively recently, ac parametric drives have been predicted and shown to be able to sustain localized waves in a linear Schr\"odinger system. In the undriven case the excitations would simply disperse \cite{dunl86}. This effect is referred to as dynamic localizations. Recently, it has been suggested theoretically \cite{lenz03,long05} and shown experimentally \cite{long06,szam10} that periodically curved optical waveguide arrays can be an ideal system for realizations of dynamic localizations. Such a localization has been used in quantum physics in the context of Bose-Einstein condensates to reduce or even completely suppress quantum tunneling of particles trapped in a potential well by shaking the potential back and forth (see the review \cite{mors10}). The tunneling suppression in a series of potential wells using the method has been shown experimentally \cite{lign07,sias08,zene09}. The method has been proposed as well as a powerful tool to manage the dispersion of, e.g., an atomic wavepacket \cite{cref09} and nonlinear gap-solitons \cite{blud04}.

\begin{figure}[tbhp!]
\begin{center}{\includegraphics[width=8.5cm]{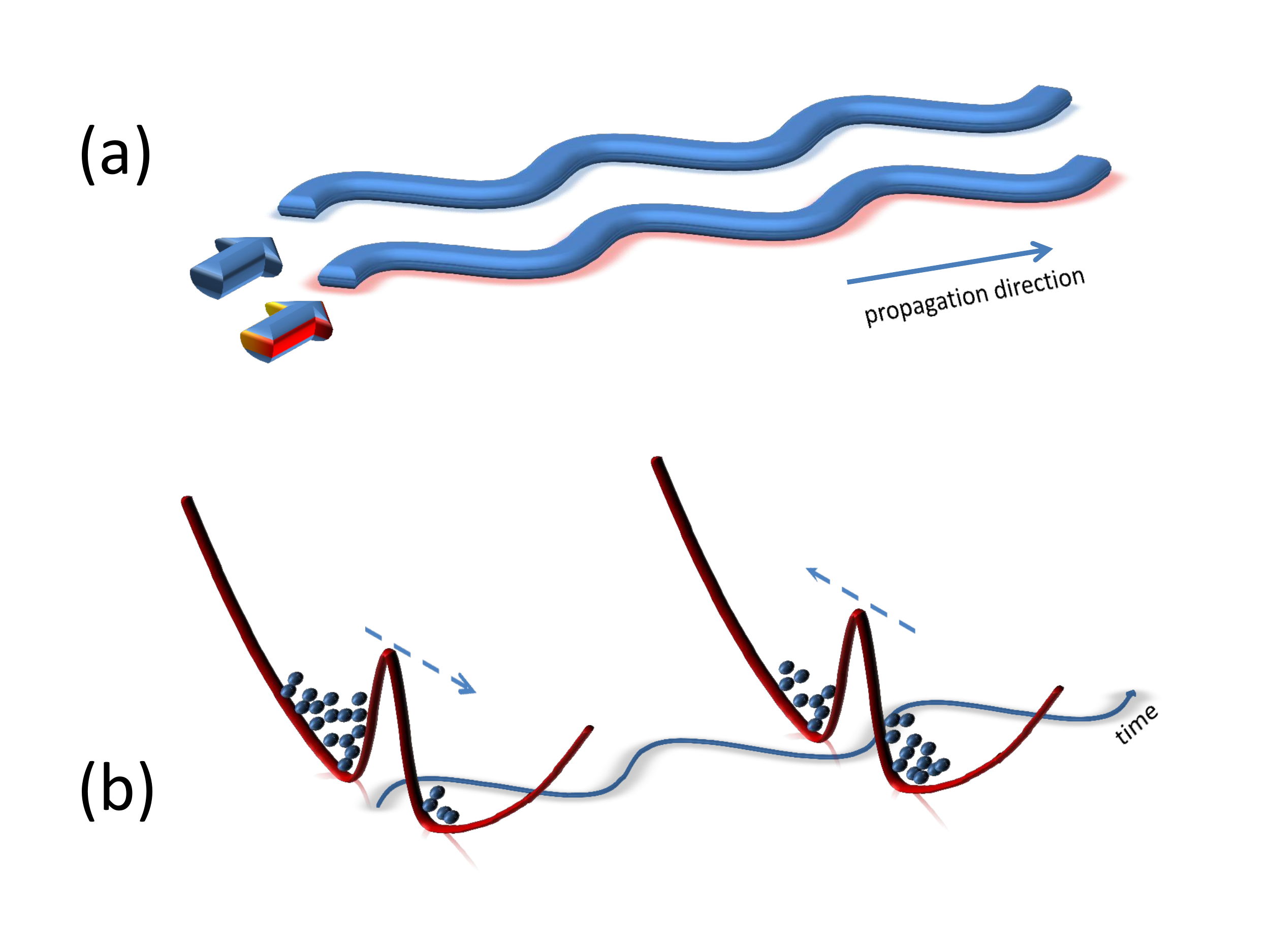}}
\end{center}
\caption{Sketch of (a) periodically-curved couplers, (b) Bose-Einstein condensates in a periodically oscillating double-well potential.}
\label{fig_0}
\end{figure}

Complementing the seminal finding of dynamic localization, coherent destruction of tunneling predicted in \cite{gros91,gros91b} has become an important phenomenon in the study of quantum dynamical control. While dynamic localization occurs in infinite dimensional systems, system boundaries play an important role in the coherent destruction of tunneling. The internal relationship between the two phenomena was discussed recently in \cite{kaya08} where it was shown that both phenomena can be interpreted as a result of destructive interference in repeated Landau-Zener level-crossings. A recent experiment on strongly driven Bose-Einstein condensates in double-well potentials formed by optical lattices provided a direct observation of dispersion suppression in matter waves \cite{kier08}. In the context of optical physics, the idea of tunneling destruction in couplers, i.e.\ dual-core waveguide arrays, was proposed in \cite{long05b} and later implemented experimentally in \cite{vall07,szam09}. A sketch of the physical setups is shown in Fig.\ \ref{fig_0}.

It is important to note that the above phenomena are in the linear regime (see a recent review on theoretical and experimental advances in modulated photonic lattices that provide controls on fundamental characteristics of light propagations \cite{gara12}). The parameter values for tunneling suppression are isolated degeneracy points of the quasienergies. When nonlinearity is present in the system, the tunneling can be suppressed in a relatively wide interval of parameter values \cite{luo07}. It is also interesting that periodic driving may enhance tunneling as opposed to preventing it \cite{lin90,pere91}, which has been shown experimentally recently \cite{voro03}.

Despite many important works mentioned above, it is necessary to note that so far there is no study on the effects of parametric drives from a dynamical system point of view. In that regard, the present work aims at studying the effects of nonlinearity on the aforementioned tunneling enhancements or suppression. In particular, we consider the persistence, i.e.\ the existence and stability, of equilibrium states when the potential is periodically oscillating. As discussed in the theoretical \cite{smer97,ragh99} and experimental \cite{albi05,levy07,zibo10} studies of Josephson tunneling of Bose-Einstein condensates in a double-well potential, the dynamics of the macroscopic quantum density and phase difference between two condensates is described by two coupled nonlinear ordinary differential equations. Depending on a control parameter, one can obtain qualitatively different dynamical behaviors that can be explained well by classical bifurcation theory. In general, there are three types of fixed-points in the phase-portraits. Using notations in the limit of no coupling between the condensates, the equilibria can be symbolically written as $[+,+]$, $[+,0]$, and $[+,-]$ where "$+$" and "$-$" represent 0 and $\pi$-phase of the wavefunctions in the two potential minima, respectively, and "0" the case when the density vanishes in the well. The second equilibrium state is also commonly referred to as a self-trapped state because the density difference between the wells oscillates about a nonzero value. The effects of parametric drives, i.e.\ an oscillating potential, on such equilibria are studied numerically in the present work. It will be shown that the parametric drive in general does not influence the existence of an equilibrium, which becomes a relative periodic orbit due to the drive, but indeed may alter the stability. In particular, we will show that the self-trapped states, i.e.\ no-tunneling states, can be broken by the drive creating, possibly chaotic, tunneling between the two wells, i.e.\ a destruction of self-trapping with the relative phase of the fields oscillating decoherently. In addition to that, we will also show that the symmetric $[+,+]$-state which is unstable beyond a critical norm can be made stable by the drive, which we call a \emph{coherent construction of tunneling}. 
The antisymmetric $[+,-]$-state, which is generally robust, can also be destabilized by the parametric drive.

The paper is presented as follows. In Section II, we discuss the governing equation and the numerical methods used in the paper. In the following section, the destruction of self-trapped states due to a parametric drive is presented. We show that the destruction is caused by a period-doubling bifurcation. In Section IV, we study the persistence of symmetric states. We show that parametric drives are able to stabilize or destabilize the states. A similar numerical discussion for the antisymmetric state is presented in Section V. The state which is always stable in the undriven case can become unstable due to parametric drives. Analytical calculations are presented in Section VI. We discuss the (de)stabilization intervals observed in the previous sections using an averaging method. The symmetry breaking bifurcation point where the asymmetric and symmetric states merge is shown using perturbation theory to be affected by parametric drives. Conclusions are presented in the last section.

\section{Governing equation and numerical methods}
\label{sec2}

Using a tight-binding approximation, the parametrically driven wavefunctions due to an oscillating potential, i.e.\ in periodically oscillating waveguides in  nonlinear optics or horizontally shaken double-well magnetic traps in matter waves, are described by (see, e.g., \cite{cref09,long05,szam10})
    \begin{equation}
    \displaystyle
    \begin{array}{lll}
    i\dot{u}_1 = \delta\left|u_1\right|^2u_1-q\,u_1+c\,e^{-ix_0(z)}u_{2},\\
    i\dot{u}_2 = \delta\left|u_2\right|^2u_2-q\,u_2+c\,e^{ix_0(z)}u_{1}.
    \end{array}
    \label{gov}
    \end{equation}
In the context of nonlinear optics, $u_j$ is the optical field in the $j$th waveguide, $c>0$ is the waveguide coupling coefficient (in units of $1/$mm), the dot is a derivative with respect to the propagation direction $z$, $q$ is the light propagation constant and $\delta>0$ is the nonlinearity coefficient ($1/$(W$\cdot$mm)). The defocusing case $\delta<0$ can be obtained immediately due to the staggering transformation $u_j\to(-1)^ju_j$. The parametric drive is represented by the function $x_0(z)=(n_s\alpha /h) \dot{\tilde{x}}_0(z)$, where $\alpha$ is the separation distance between the waveguides in units of $\mu$m, $n_s$ is the substrate refractive index, $\tilde{x}_0(z)$ describes the physical periodic curving profile and $h$ is the inverse of the light wavenumber. In this work, we take ${x}_0(z)=a\omega\sin(\omega z)$ with amplitude $a$ in units of mm and waveguide curvature wavenumber $\omega$. The constant, time-independent quantity $N=|u_1|^2+|u_2|^2$ associated with the field power is referred to as a norm herein. The parameters are taken in the vicinities of those used in \cite{luo07}. The onset of, e.g., the existence and the stability of a solution will certainly depend on the parameter values, nevertheless the results presented herein are qualitatively generic. Without loss of generality, one can scale $\delta=\omega=1$. When $a=0$, Eq.\ (\ref{gov}) is integrable.

To study the persistence of an equilibrium in the presence of a parametric drive, we solve the governing equation (\ref{gov}) for periodic orbits. We therefore seek solutions satisfying $u_n(0)=u_n(T)$, where $T=2\pi$ is the oscillation period. In the absence of a drive, equilibrium solutions of the equations clearly fulfil the relation. In the presence of drives, we look for the continuations of an equilibrium point by using shooting methods in real space or algebraic methods by discretizing the propagation direction variable. The latter method is more useful than the first when there is a bifurcation in the numerical continuation. In that case, we use a pseudo-arclength continuation algorithm \cite{gova00,kell87,kuzn98}. 

When a periodic orbit, say $U_n(z)$, is obtained, we also examine its stability by calculating its Floquet multipliers, which are eigenvalues of the monodromy matrix. This is obtained from solving a linearized equation about the solution $U_n(t)$
    \begin{equation}
    \displaystyle
    \begin{array}{lll}
    i\dot{u}_1 = \delta\left(2\left|U_1\right|^2u_1+U_1^2u_1^*\right)-q\,u_1+c\,e^{-ix_0(z)}u_{2},\\
    i\dot{u}_2 = \delta\left(2\left|U_2\right|^2u_2+U_2^2u_2^*\right)-q\,u_2+c\,e^{ix_0(z)}u_{1}.
    \end{array}
    \label{lin-gov}
    \end{equation}
The linear system is integrated using a Runge-Kutta method of order four or a symplectic method. The $i$th column of the monodromy matrix $M$ is the vector $[\textrm{Re}(u_1(T)),\textrm{Re}(u_2(T)),\textrm{Im}(u_1(T)),\textrm{Im}(u_2(T))]^\mathrm{T}$, that corresponds to the initial condition $[\textrm{Re}(u_1(0)),\textrm{Re}(u_2(0)),\textrm{Im}(u_1(0)),\textrm{Im}(u_2(0))]^\mathrm{T}$ that is equal to the $i$th column vector of the identity matrix $I_{4}$. A solution is stable when all the multipliers are on the unit circle. Note that here it is not necessary to have a pair of eigenvalues at $(+1)$. The (in)stability result obtained from calculating the monodromy matrix is also confirmed by integrating Eq.\ \eqref{gov}.

Writing $u_j=|u_j|e^{i\phi_j z}$, it is convenient to represent a solution with its population imbalance $\Delta$ and phase-difference $\theta$ between the light fields described as
\begin{equation}
\Delta=(|u_1|^2-|u_2|^2)/N,\quad\theta=\phi_2-\phi_1.
\end{equation}
One can show that $\Delta$ and $\theta$ satisfy parametrically driven sine-Gordon equations
    \begin{equation}
    \begin{array}{lll}
    \dot{\Delta} = 2\,c\sqrt {1-{\Delta}^{2}} \left(\sin {x_0} \cos {
\theta} -\cos {x_0} \sin{\theta} \right),
 \\
    \dot{\theta} = \frac {2\Delta c}{\sqrt {1-{\Delta}^{2}}}\left(\cos {x_0} \cos  {\theta} + \sin {x_0}  \sin {\theta} \right)-\Delta N\delta.
    \end{array}
    \label{dth}
    \end{equation}
When there is no drive, i.e.\ $x_0\equiv0$, we obtain the equations derived in \cite{smer97}. 
In that case, (\ref{dth}) will have at most three fixed points given by $(\theta,\Delta)=(0,0)$, $(\pm\pi,0)$, and $(0,\pm\sqrt{1-\left(2c/(\delta N)\right)^2})$, which correspond respectively to the symmetric, antisymmetric, and asymmetric state. It is clear that the asymmetric state pair only exists when $c<\delta N/2$. The symmetric and antisymmetric states in the focusing case correspond respectively to the antisymmetric and symmetric states of the defocusing case due to the staggering transformation.

\section{Destabilization of asymmetric self-trapped states}

\begin{figure}[tbhp!]
\begin{center}
\subfigure[]{\includegraphics[width=7cm]{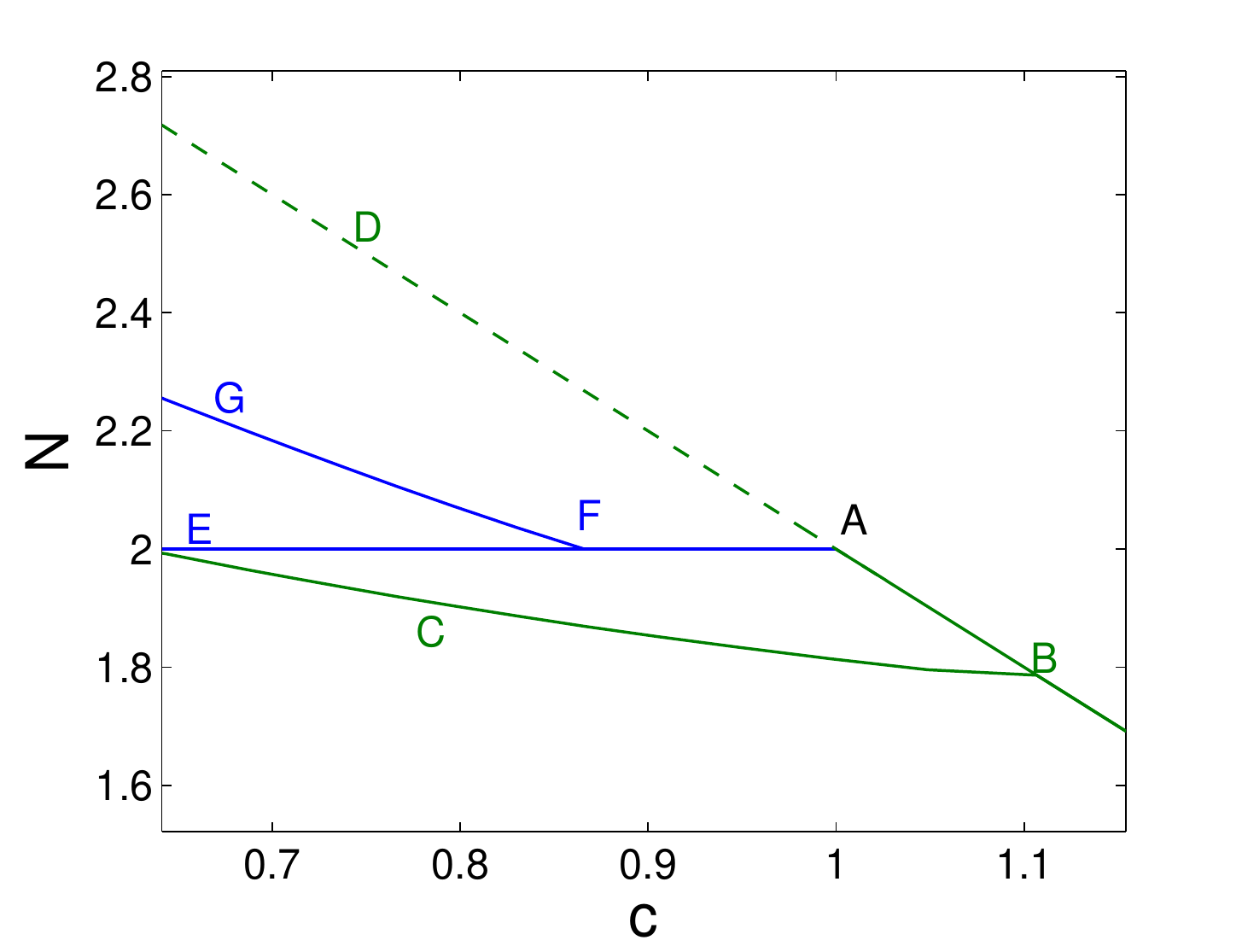}}\\
\subfigure[]{\includegraphics[width=7cm]{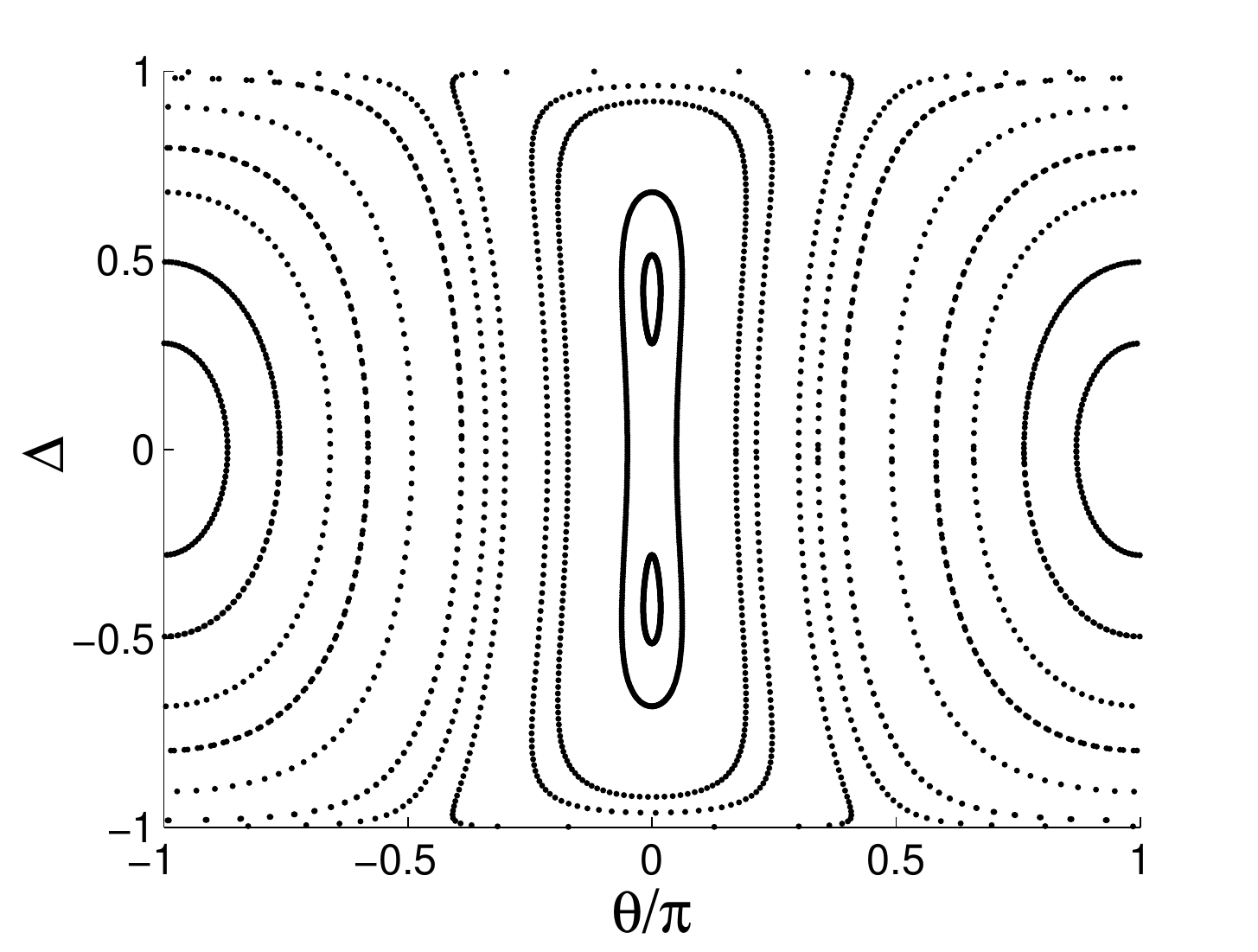}}
\end{center}
\caption{(a) Bifurcation diagram of equilibrium states in straight couplers $a=0$. Shown is the norm of the states against the coupling parameter $c$ with $q=2$. Dashed line indicates unstable solutions. (b) Phase-portraits of (\ref{gov}) with $a=0,\,N=2$ and $c=10/11$.}
\label{fig_1}
\end{figure}

First, we consider the effect of a parametric drive on the asymmetric ground state.  Let us set the propagation constant, e.g., $q=2$. When $a=0$, 
the bifurcation diagram of the stationary $[+,+]$ state is shown in Fig.\ \ref{fig_1}(a).

The symmetric $[+,+]$-state corresponds to branch $BD$. For a sufficiently small norm, the state is known to be stable (see, e.g., \cite{kirr08,kirr11} and references therein). When one decreases the parameter $c$ further, there will be a critical norm above which the symmetric state is no longer stable. The equilibrium loses stability at a pitchfork (symmetry breaking) bifurcation with an asymmetric state leading to "macroscopically quantum self-trapping" in the context of matter waves \cite{smer97,ragh99}, i.e.\ point $A$ in the figure. The state corresponding to branch $AE$ can be viewed as 
the $[+,0]$-state. In Fig.\ \ref{fig_1}(b), we show the phase-portraits of the coupled equations (\ref{gov}) in the $(\theta,\Delta)$-plane with $N=2$ and $c=10/11$. One can see that the symmetric state represented by $(0,0)$ is indeed unstable while there is a pair of stable fixed points lying on the vertical axis $\theta=0$, which is the asymmetric ground state. The $BC$ branch in Fig.\ \ref{fig_1}(a) corresponds to a periodic orbit encircling both the stable fixed points with $\theta=0$ in Fig.\ \ref{fig_1}(b). The branch $FG$ corresponds to a periodic orbit encircling only one of the stable fixed points. Both branches do not correspond to equilibria, but are shown for the completeness of the analysis later. 

Next, we consider the presence of a parametric drive in the system. Rather than showing the phase portraits in the $(\theta,\Delta)$-plane with a continuous time, it becomes more convenient to represent the solution trajectories in Poincar\'e maps (stroboscopic plots at every period $T=2\pi$). Note that in a Poincar\'e map a periodic orbit will correspond to a fixed point. 
In the recurrence map, a stable periodic orbit will therefore correspond to an elliptical fixed point encircled by closed regions (islands).

\begin{figure}[tb!]
\begin{center}
\subfigure[]{\includegraphics[width=7cm]{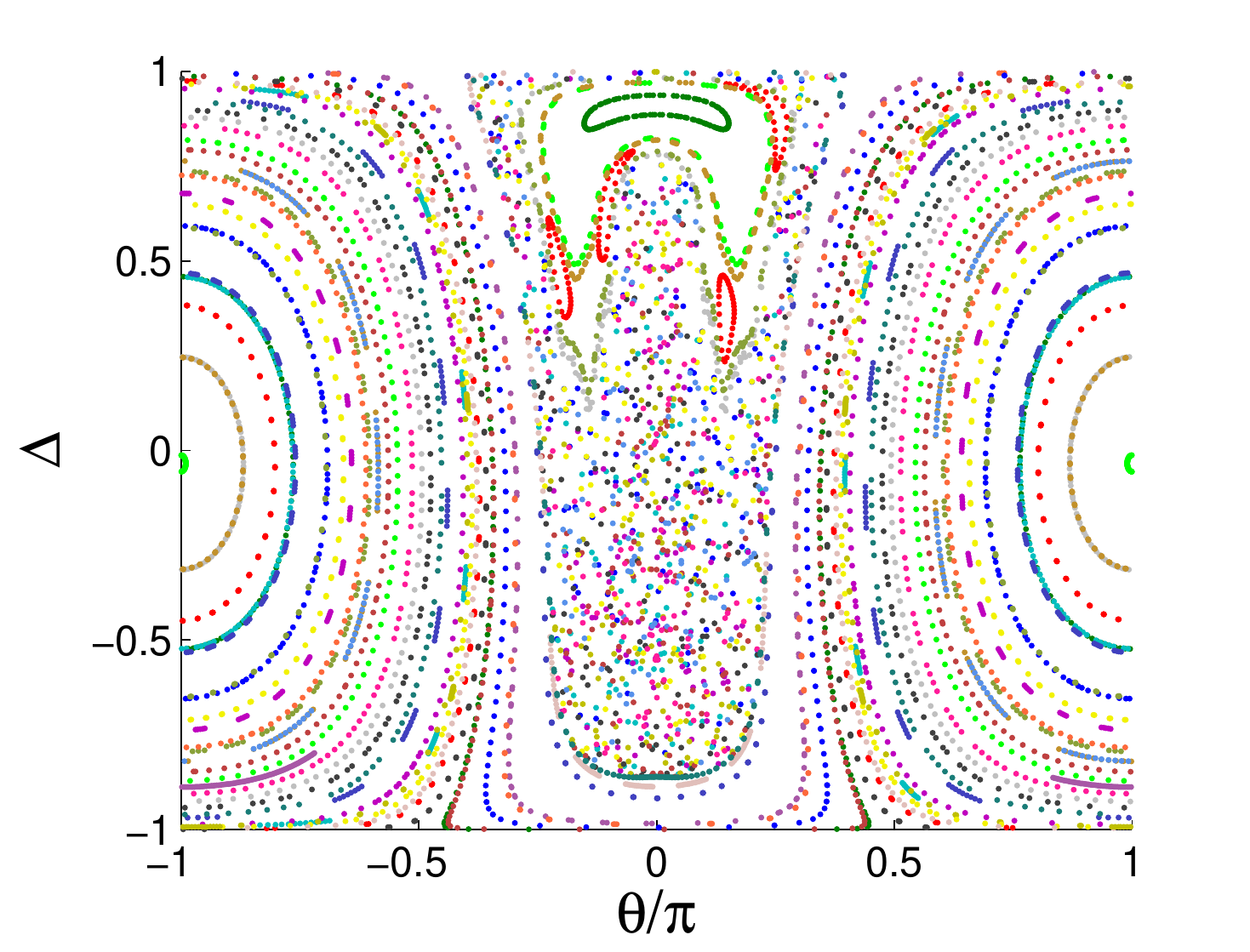}}\\
\subfigure[]{\includegraphics[width=7cm]{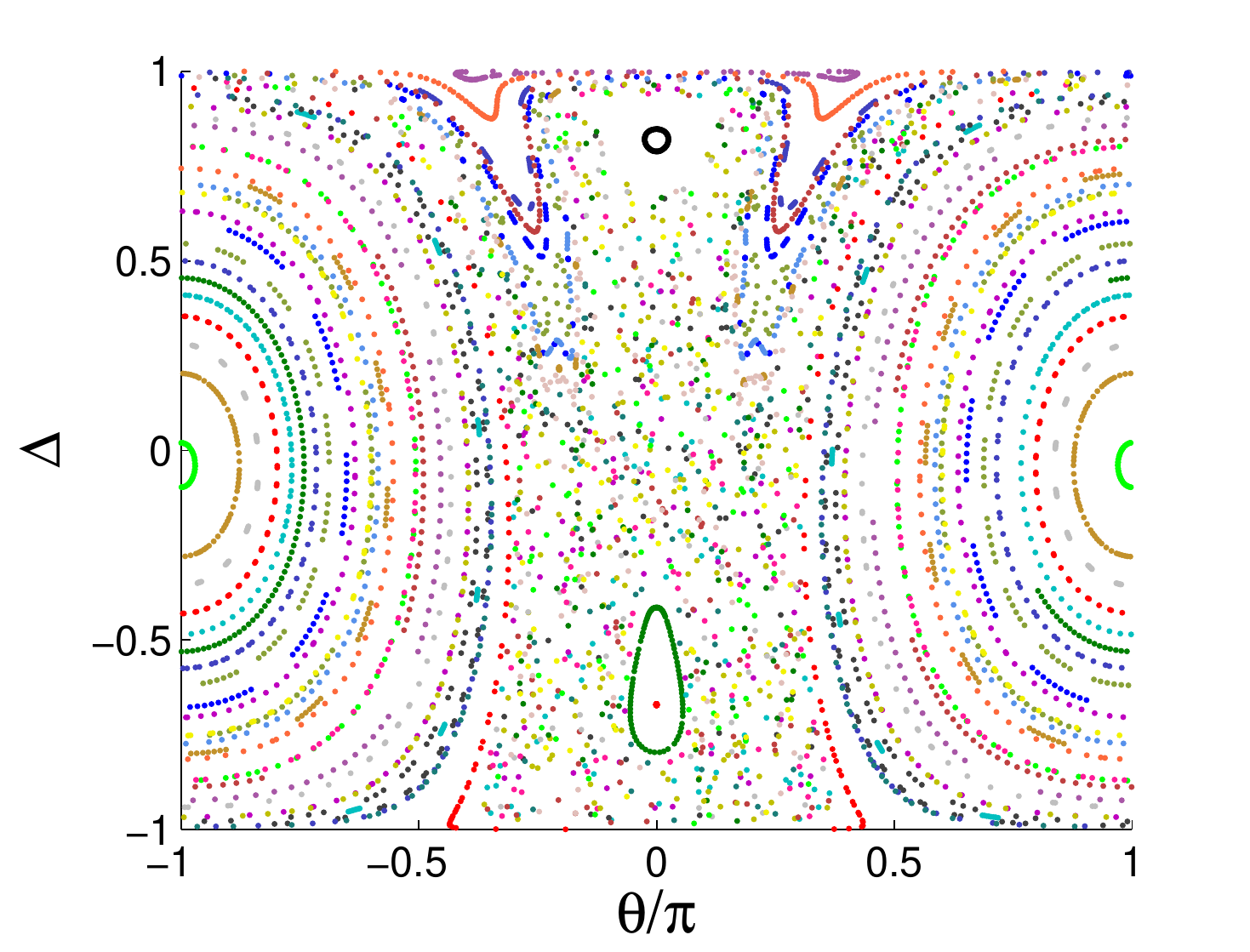}}
\end{center}
\caption{ (Color online) Stroboscopic plots of the system (\ref{gov}) with $N=2$ and (a) $c=10/11$ and (b) $c=2/3$. 
Dots with the same color are obtained from the same initial condition.}
\label{fig_2}
\end{figure}

\begin{figure}[tb!]
\begin{center}
{\includegraphics[width=7cm]{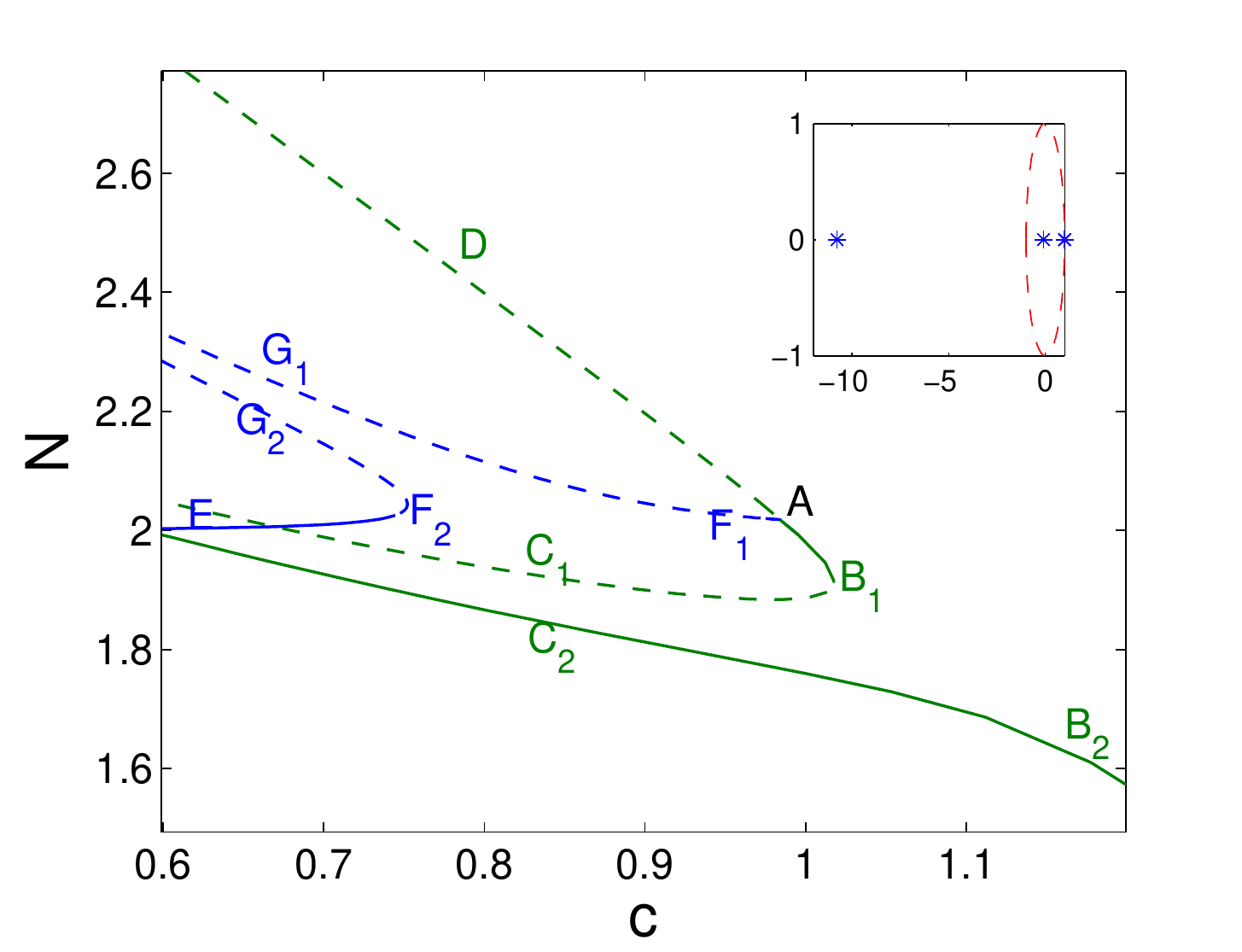}}
\end{center}
\caption{Bifurcation diagram of equilibrium states in periodically curved couplers with $a=0.1$. 
The inset shows Floquet multipliers of the asymmetric state in the complex plane for $c=10/11$.}
\label{fig_3}
\end{figure}

Setting $a=0.1$, we show in Fig.\ \ref{fig_2} the recurrence maps of the system for norm $N=2$ with two different values of $c$, i.e.\ $c=10/11$ and $2/3$, in the $(\theta,\Delta)$-plane. The plots are obtained from various sets of initial conditions using direct numerical integrations of the governing equations (\ref{gov}). It is important to mention that in the first case, there is no asymmetric trapped state. The island with $\theta=0$ in the figure is actually a tunneling state ($\Delta(t)$ is not sign-definite), that is locked to and oscillates coherently with the drive. It is also clear that there are rather chaotic oscillations between the fields in the two potential minima. Interestingly, by changing 
the coupling constant, one now has self-trapped states shown in Fig.\ \ref{fig_2}(b) by a pair of small islands centered at $\theta=0$. The question is: what happens with the states in the first case? 
We 
conjecture that the two topologically different maps in Fig.\ \ref{fig_2} are determined by the persistence of such fixed points. To illustrate our conjecture, we have solved the governing equations (\ref{gov}) for periodic orbits. The corresponding bifurcation diagram of Fig.\ \ref{fig_1}(a) when $a=0.1$ is shown in Fig.\ \ref{fig_3} 
where a rich bifurcation structure is evident.

Comparing Figs.\ \ref{fig_1}(a) and \ref{fig_3}, one can see that there is also branch merging and splitting. Branches $FG$ and $BC$ in Fig.\ \ref{fig_1}(a) split into $F_jG_j$ and $B_jC_j$, $j=1,2$ in Fig.\ \ref{fig_3}. More importantly, the branch $F_1G_1$ merges with a segment $AF$ that corresponds to an asymmetric ground state, which now becomes unstable. The norm used in Fig.\ \ref{fig_2}(a) belongs to this unstable region, explaining why there is no asymmetric trapped state in the Poincar\'e section. In the inset of Fig.\ \ref{fig_3}, we display the Floquet multipliers of the state showing that it suffers from an instability due to a pair of multipliers leaving at $-1$, i.e.\ a period-doubling bifurcation. The question whether subsequent period-doubling bifurcations occur in the system is left for further future investigations.

The presence of only one family of islands in Fig.\ \ref{fig_2}(a) centered at $\theta=0$, i.e.\ without its counterparts in the lower half-plane, can also be explained using the bifurcation diagram in Fig.\ \ref{fig_3}. When $a=0$, one of the islands corresponds to a periodic orbit along the branch $BC$. As the branch splits into unstable $B_1C_1$ and stable $B_2C_2$ branches, it is expected that there will be only one family of stable coherent periodic orbits as seen in the Poincar\'e sections. The splitting itself is expected from the Floquet multipliers of the $BC$ branch as they are all at $+1$ (not shown here). This is also the case with branch $FG$.

The destabilization of the asymmetric state is rather generic, including the case of small $N$. One difference is that we only observed unstable solutions and did not see any branch splitting along the asymmetric state branch unlike in Fig.\ \ref{fig_3}.

\section{Persistence of symmetric states}

\begin{figure}[tb]
\begin{center}
\subfigure[]{\includegraphics[width=7cm]{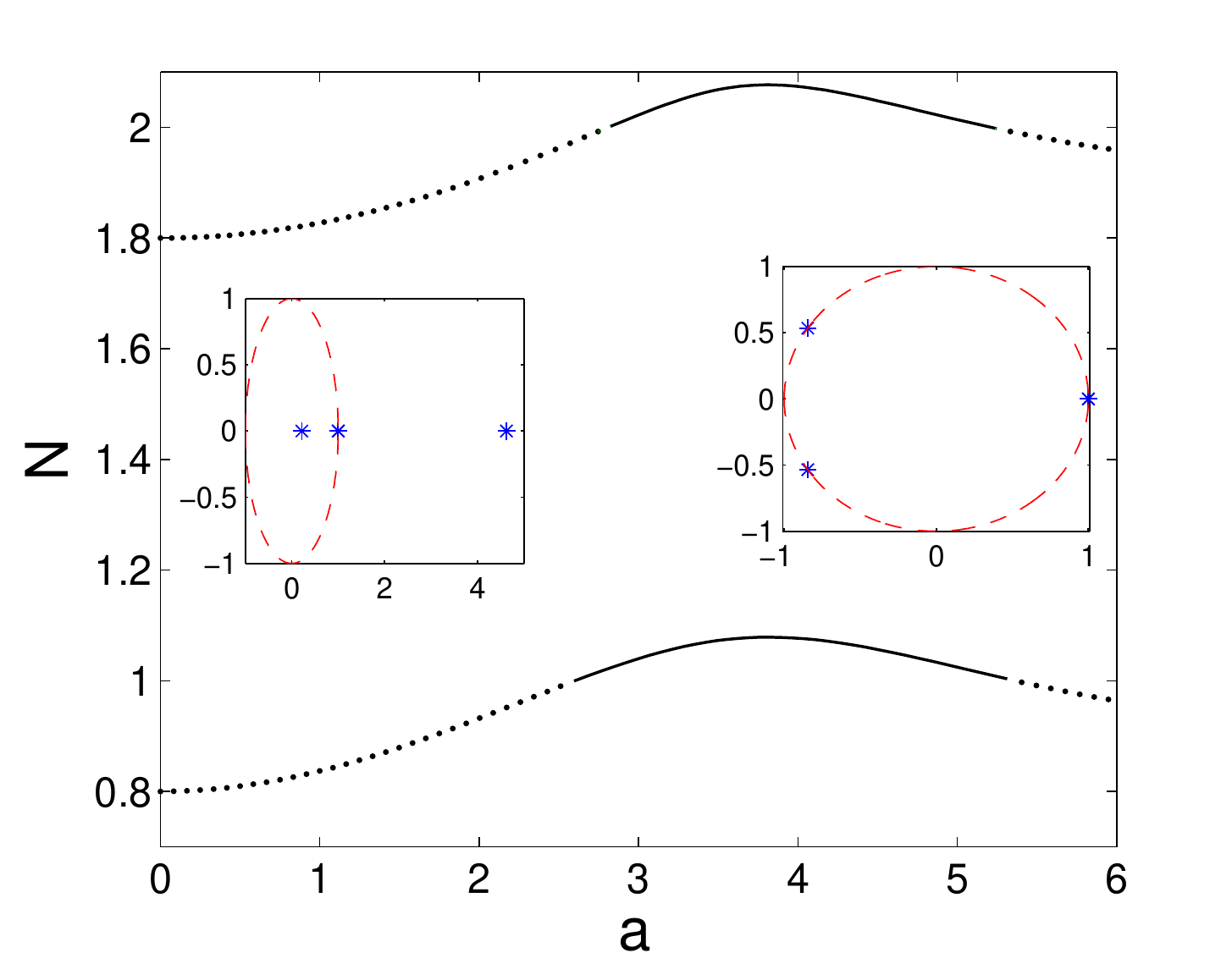}}
\subfigure[]{\includegraphics[width=7cm]
{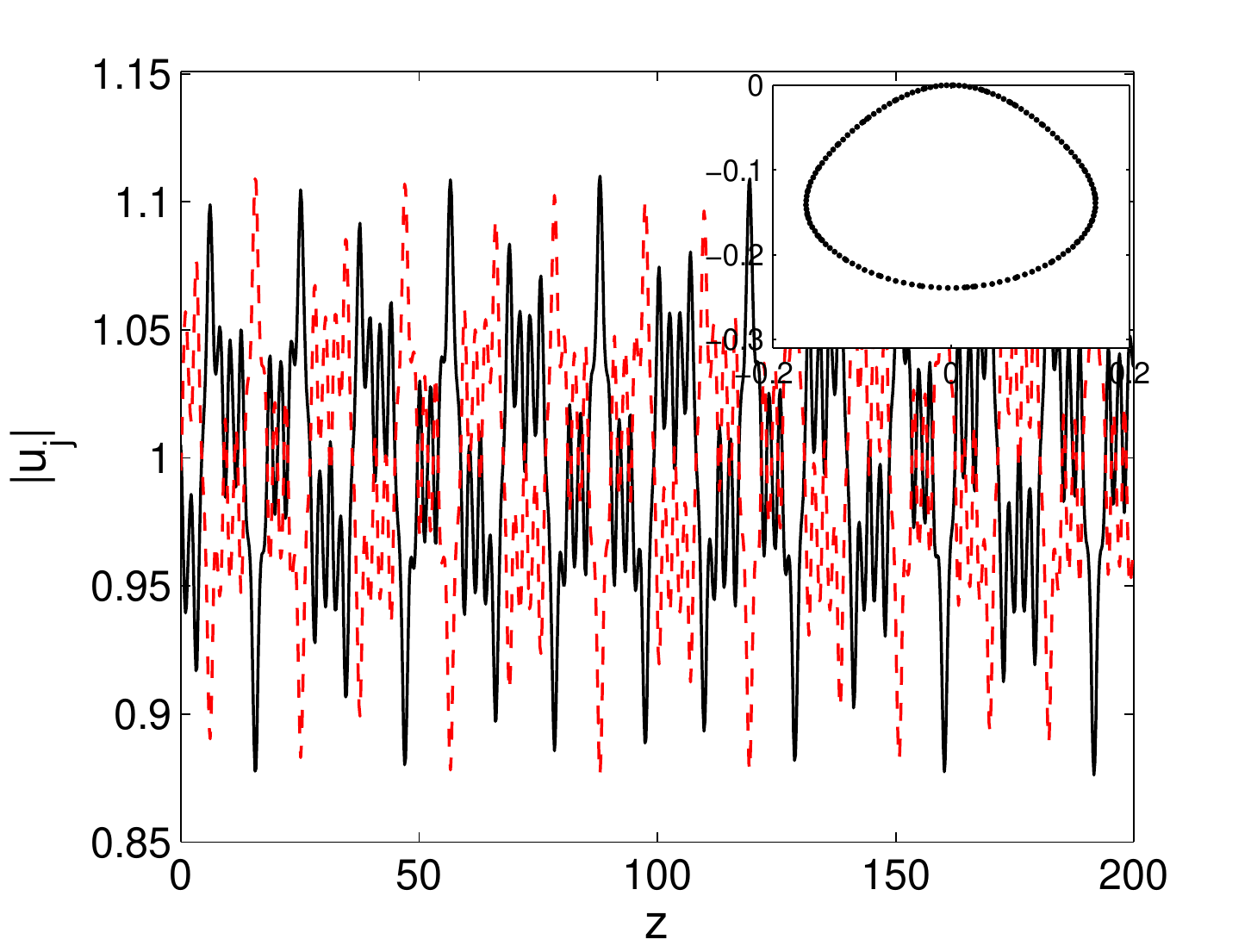}}
\end{center}
\caption{(a) The norm of symmetric states as a function of the amplitude drive $a$ for $c=0.1$ with $q=2$ (upper) and $q=1$ (lower). Dotted line indicates unstable solutions. The insets show the Floquet multipliers of two solutions along the upper curve at $a=2.5$ and $a=3.5$. (b) The time dynamics of the initial condition $u_1=u_2$ 
with $c=0.1,\,N=2$ and $a=3.5$. The inset 
depicts the trajectories of the dynamics in time in the $(\theta,\Delta)$-plane.}
\label{fig_5}
\end{figure}

We have seen in Section \ref{sec2} that in the undriven case when the solution norm exceeds a critical value, the symmetric states lose stability to asymmetric states, see Figs.\ \ref{fig_1}. 
It is natural to question whether it is possible to stabilize unstable symmetric states using the parametric drive. We show in Fig.\ \ref{fig_5} that it is indeed the case.

In Fig.\ \ref{fig_5}(a) we depict the norm of symmetric states as a function of the drive amplitude $a$ for two values of $q$. As a particular choice for clarity, we consider a symmetric state with a relatively large norm and small coupling such that when there is no drive the wave fields are weakly coupled and rather strongly unstable. It is interesting to note that upon increasing the parameter $a$, the continuation of the undriven symmetric state becomes stable. The stabilization is generic as we also observe the same phenomenon for other values of $q$ with the stabilization threshold almost independent of $q$ as shown in the lower curve in Fig.\ \ref{fig_5}(a). In the inset of the figure, we show the Floquet multipliers of the states at $a=2.5$ and $a=3.5$ corresponding to values before and after the stabilization threshold. For the second case, the periodic orbit is stable shown by the multipliers that are all on the unit circle.

To see whether the stabilization observed in the continuation of the periodic orbits above can be viewed from direct numerical integrations of the governing equations (\ref{gov}), we show in Fig.\ \ref{fig_5}(b) the 
time dynamics of the initial condition $u_1=u_2$ at $z=0$ showing 
a symmetric state with coherent relative phase between the fields. Hence, we have obtained a coherent construction of tunneling in the region when symmetric (in-phase) states are unstable.

\begin{figure}[tbhp!]
\begin{center}
\subfigure[]{\includegraphics[width=7cm]{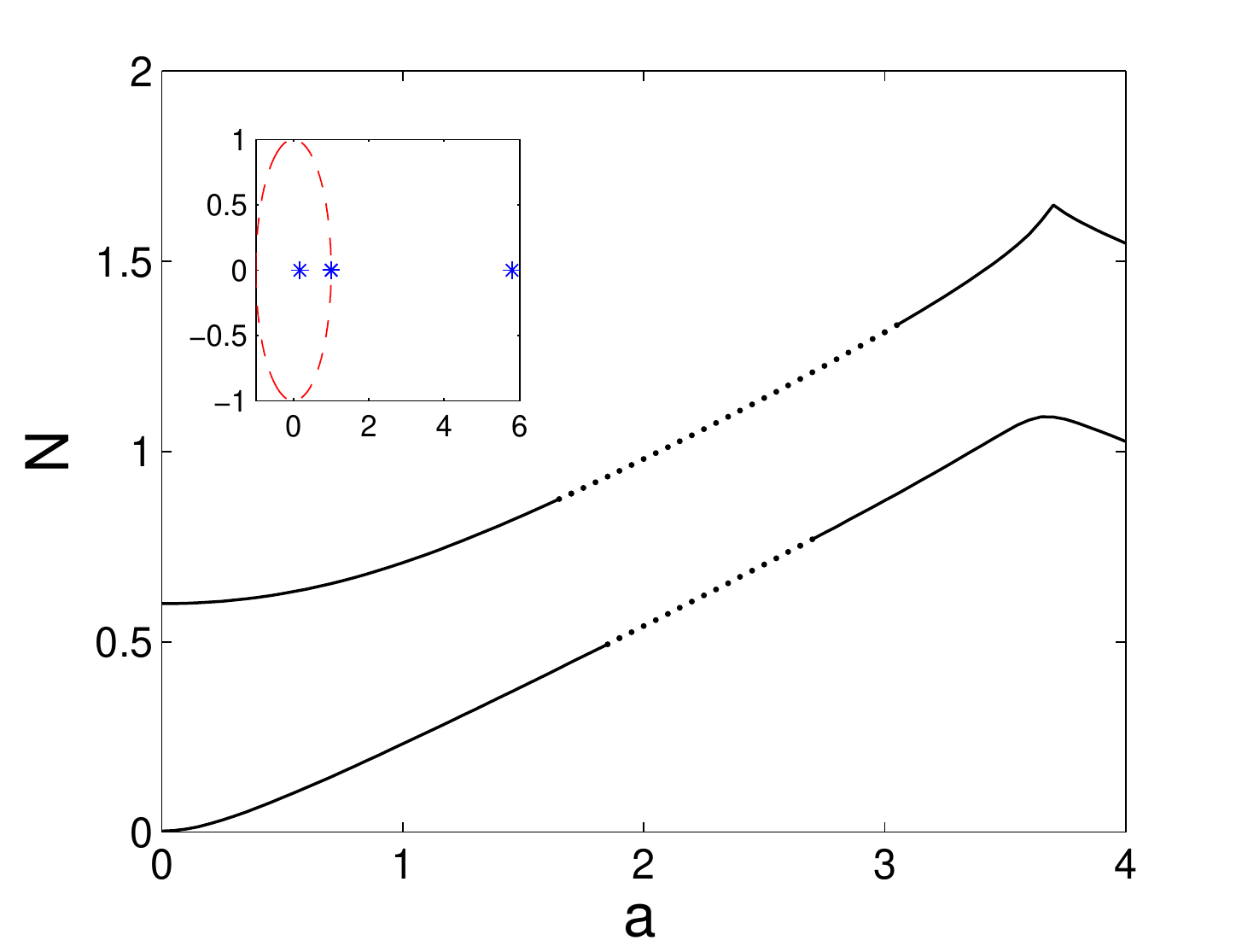}}
\subfigure[]{\includegraphics[width=7cm]{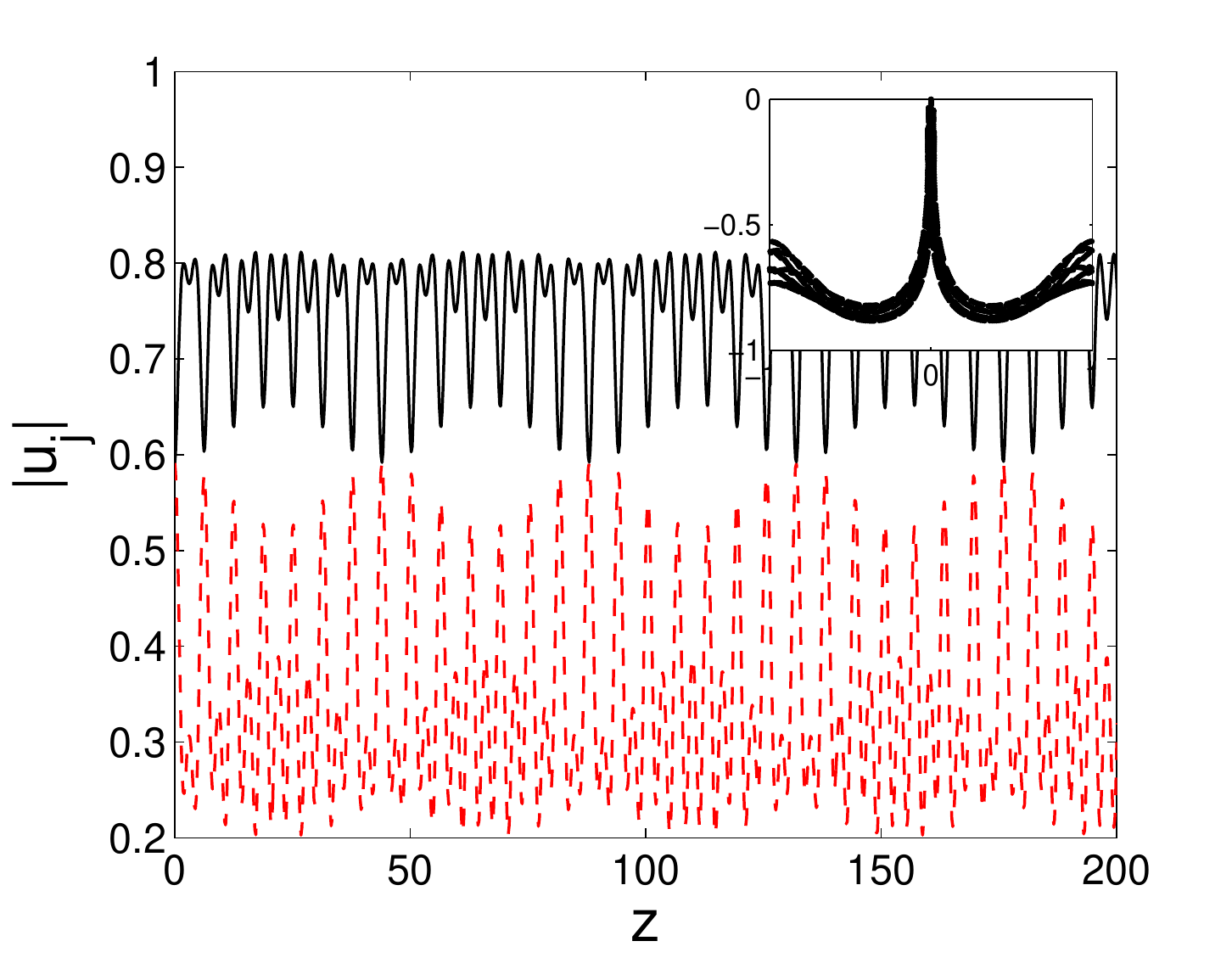}}
\end{center}
\caption{(a) The same as Fig.\ \ref{fig_5}(a), but for stable symmetric states (when undriven) with $c=0.4$. The inset shows the Floquet multipliers of a solution along the upper curve at $a=2.5$. (b) Typical dynamics of unstable symmetric states with 
$a=2$, 
$N=0.7$ and $u_1(0)=u_2(0)$. The inset shows the dynamics in the $(\theta,\Delta)$-plane.}
\label{fig_5a}
\end{figure}

In addition to stabilization, we have also studied the possibility of using a parametric drive to destabilize symmetric states that are stable in the undriven case. Our study is summarized in Fig.\ \ref{fig_5a}(a) for $c=0.4$, where we show that the continuation of an undriven symmetric state becomes unstable in an interval of driving amplitude. As for the dynamics of the instability, in Fig.\ \ref{fig_5a}(b) we depict a possible manifestation of the destabilization for $u_1(0)=u_2(0)$ and $N=0.7$.  
For this value of power, the state is stable when undriven as it is below the critical norm for a pitchfork bifurcation. 
When the state is driven, the relative phase  
is no longer localized and 
the tunneling is completely suppressed, which 
can be called a nonlinear coherent destruction of tunneling similar to that discussed in \cite{luo07}. Note that the instability window observed in Fig.\ \ref{fig_5a}(a) is relatively wide. We believe that the finite range for the nonlinear coherent destruction of tunneling reported in \cite{luo07} is due to an instability of the periodic orbits. 

\section{Destabilization of antisymmetric states}

\begin{figure}[tbhp!]
\begin{center}
\subfigure[]{\includegraphics[width=7cm]{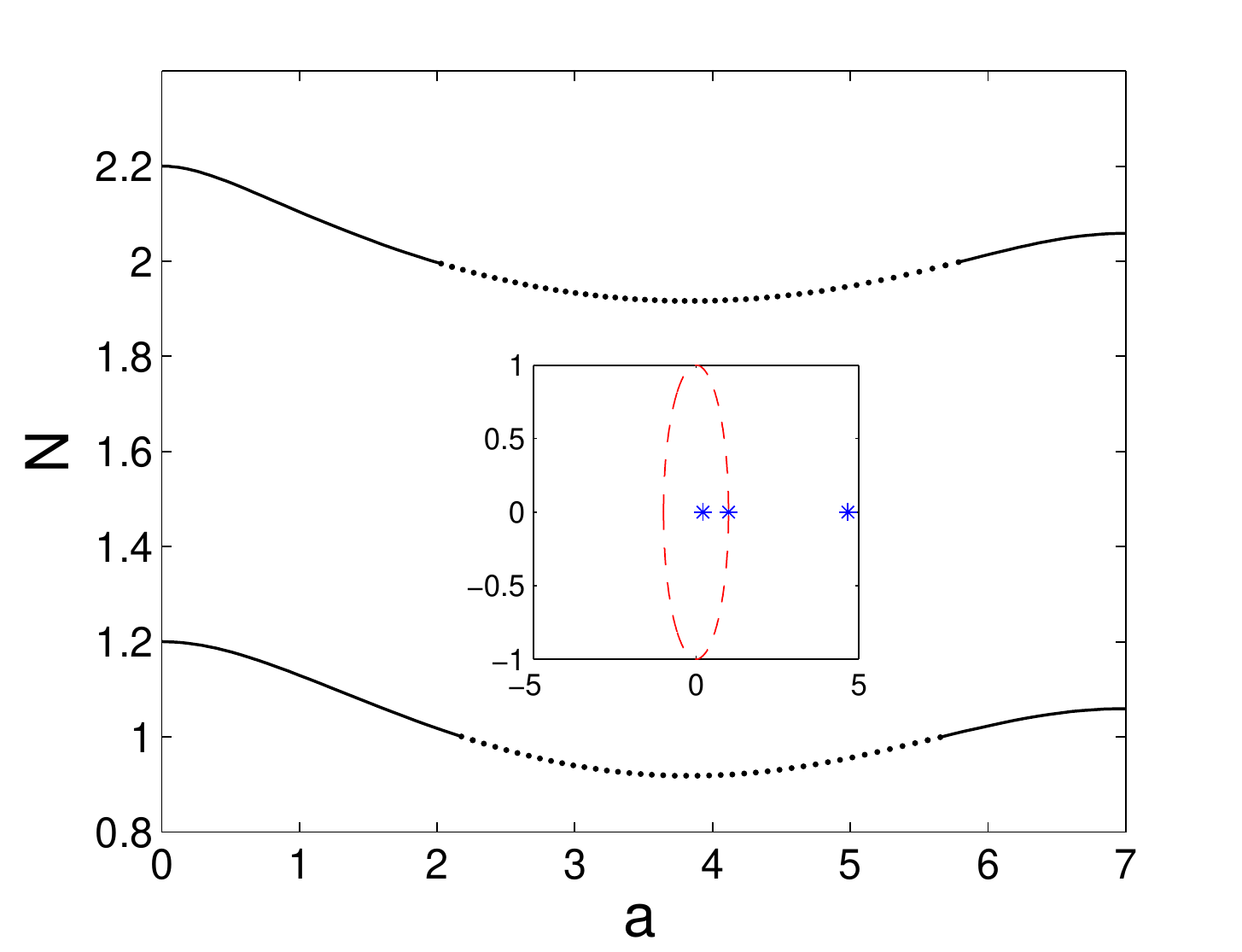}}
\subfigure[]{\includegraphics[width=7cm]{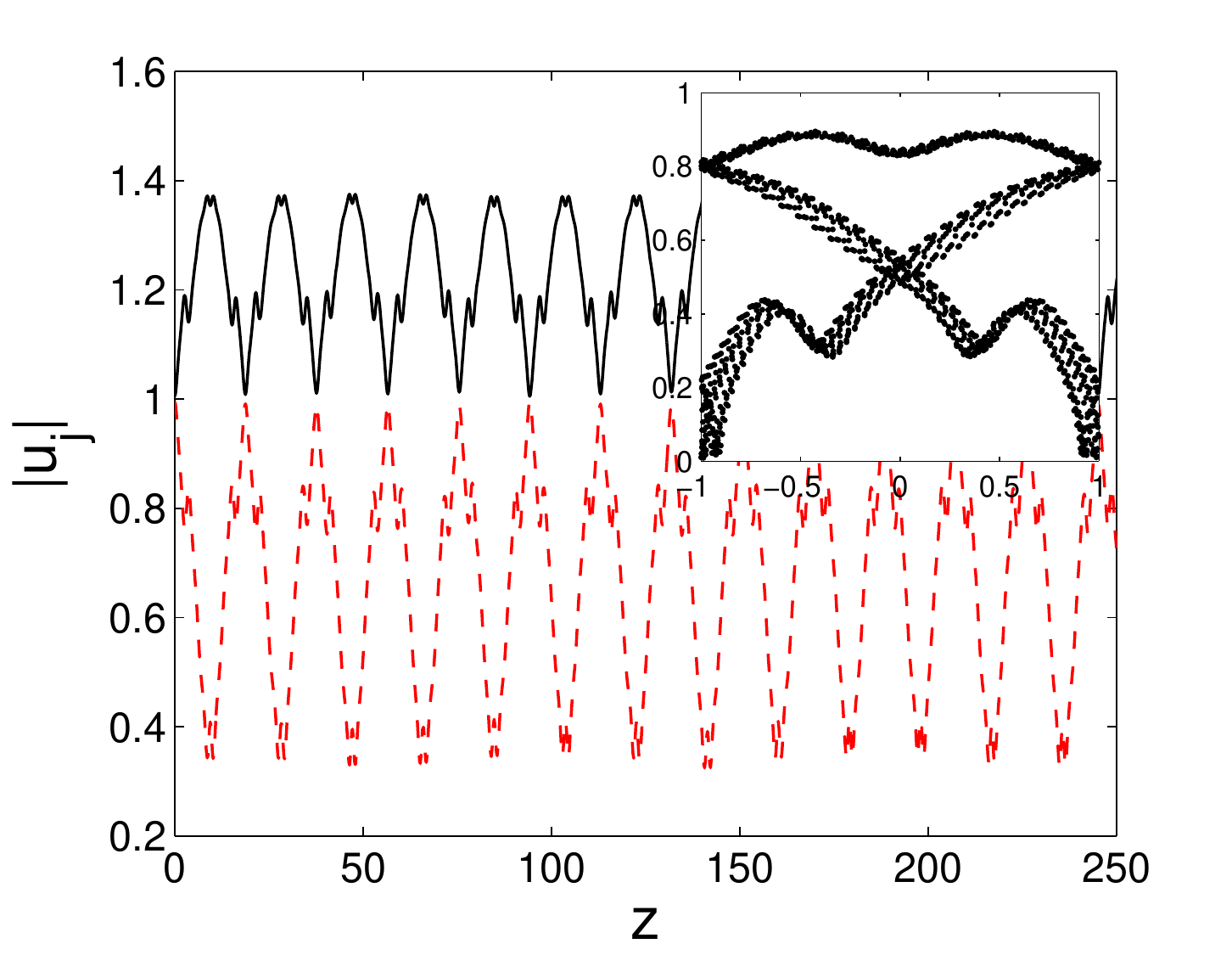}}
\end{center}
\caption{(a) The same as Fig.\ \ref{fig_5}(a), but for the antisymmetric state with $c=0.1$. The inset shows the Floquet multipliers of a solution along the upper curve at $a=2.5$. (b) The same as Fig.\ \ref{fig_5a}(b) for the antisymmetric mode with $N=2$ and $a=2.5$.}
\label{fig_6}
\end{figure}

Finally, we consider the antisymmetric $[+,-]$-state, which is stable in the undriven case. The state is generally robust. Nevertheless, in a similar fashion to the symmetric state in the previous section, we show in Fig.\ \ref{fig_6} that antisymmetric states can also become unstable when driven above a threshold value. The instability domain is finite similar to the symmetric states discussed in Section IV. When one increases the drive amplitude further, there will be other instability windows.

In Fig.\ \ref{fig_6}(b), we show the typical dynamics of unstable $[+,-]$-state, where we also obtain a nonlinear coherent destruction of tunneling similar to Fig.\ \ref{fig_5a}(b).

\section{Analytical results: stabilization, destabilization and symmetry breaking bifurcation points }

In this section we discuss the effects of parametric drives on the continuation of equilibrium states analytically.

A method commonly used in almost all of the previous theoretical studies is to average the governing equation \eqref{gov}. The approach is justified when the driving frequency is large. In our scaling, that corresponds to small norm solutions. 
The averaged equation is given by
    \begin{equation}
    \displaystyle
    i\dot{u}_j = \delta\left|u_j\right|^2u_j-q\,u_j+c J_0(a) u_{2-j},\,j=1,2,
    \label{av_gov}
    \end{equation}
where $J_0$ is the Bessel function of the first kind. Hence, we obtain a coupler with an effective coupling constant $cJ_0(a)$. 

As discussed in Section I and II above, there is a critical threshold at which the symmetric and asymmetric states change stability, i.e.\ the symmetry breaking bifurcation point $c=N/2$. Note that $J_0(a)$ oscillates about 0 with the oscillation amplitude decreasing with $a$. Therefore, in the context of the average equation the only possible way that the continuation of an equilibrium state changes stability is when the effective coupling crosses $\pm N/2$.

For the asymmetric state discussed in Section III, it is expected that the state becomes unstable in the interval of $a$ where $cJ_0(a)>N/2$. For the symmetric state studied in Section IV, the state would become stable when $cJ_0(a)>N/2$ or $cJ_0(a)<0$. The latter is due to the staggering transformation as the state effectively becomes an antisymmetric one when $cJ_0(a)$ changes sign. Finally for the asymmetric state analyzed in Section V, the state is expected to become unstable when $0>cJ_0(a)>-N/2$ due to 
the staggering transformation. The analytical explanations of the stability switching above are rather in good agreement with the numerical results.

Despite the agreement, the averaged equation fails in describing several phenomena behind the stabilization and destabilization of the original governing equation, such as the period-doubling bifurcation that causes the destabilization of the asymmetric state instead of the state ceasing to exist and the branch splitting. Another failure of the averaged equation is in predicting the influence of parametric drives on the symmetry breaking (pitchfork) bifurcation point. Because $J_0(a)<1$ for $a>0$, at the bifurcation point of the undriven case, i.e.\ $c=N/2$, the effective coupling $cJ_0(a)$ is less than half of the solution norm, which implies that the symmetric state should be unstable when the system is parametrically driven. Nevertheless, from Fig.\ \ref{fig_3} we obtained that the bifurcation point occurs at a larger norm, which contradicts the averaged equation. To resolve the issue, we used a perturbation expansion. 

\begin{figure*}[tbh]
\begin{center}
\subfigure[]{\includegraphics[width=8cm]{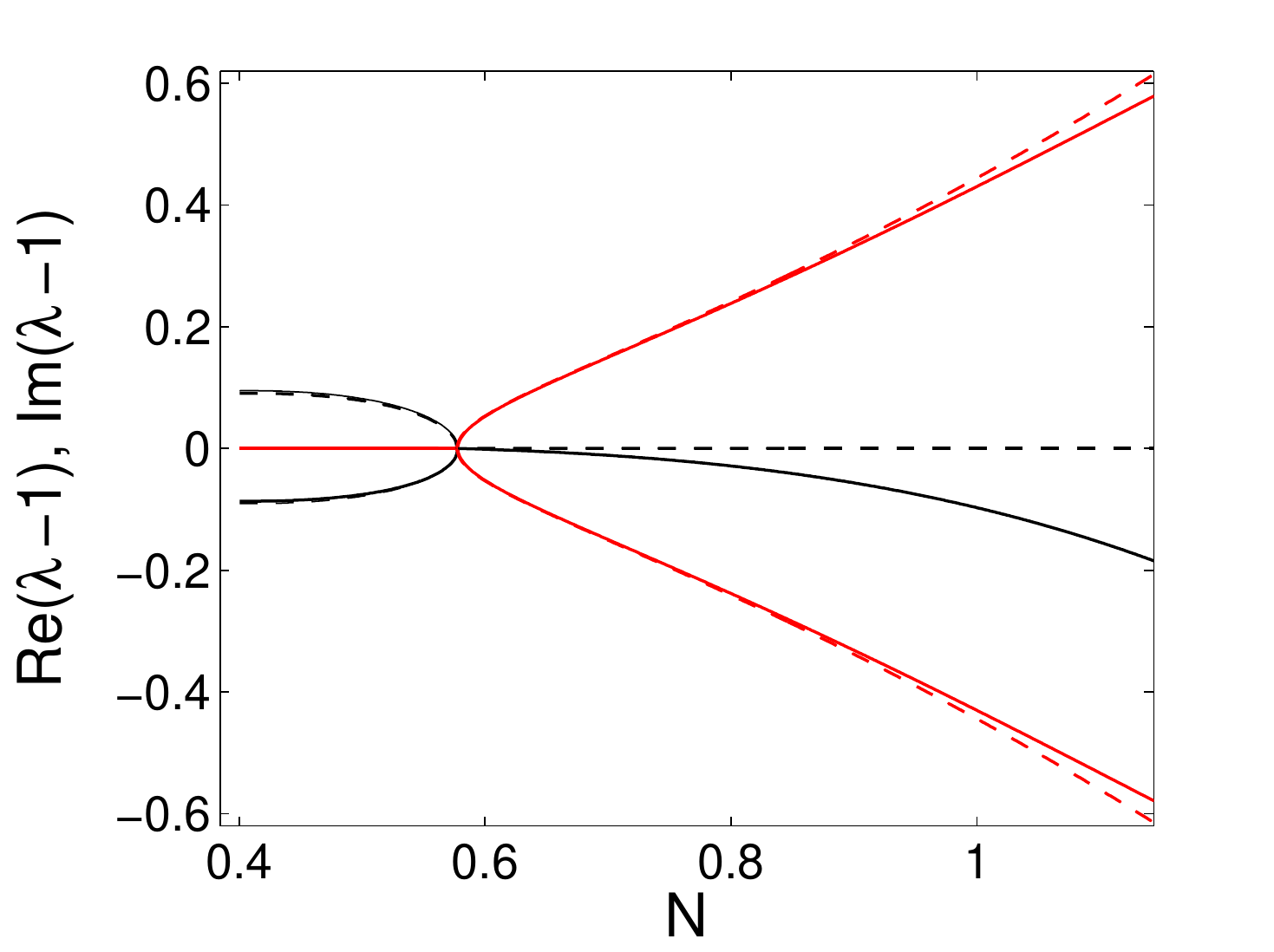}}
\subfigure[]{\includegraphics[width=8cm]{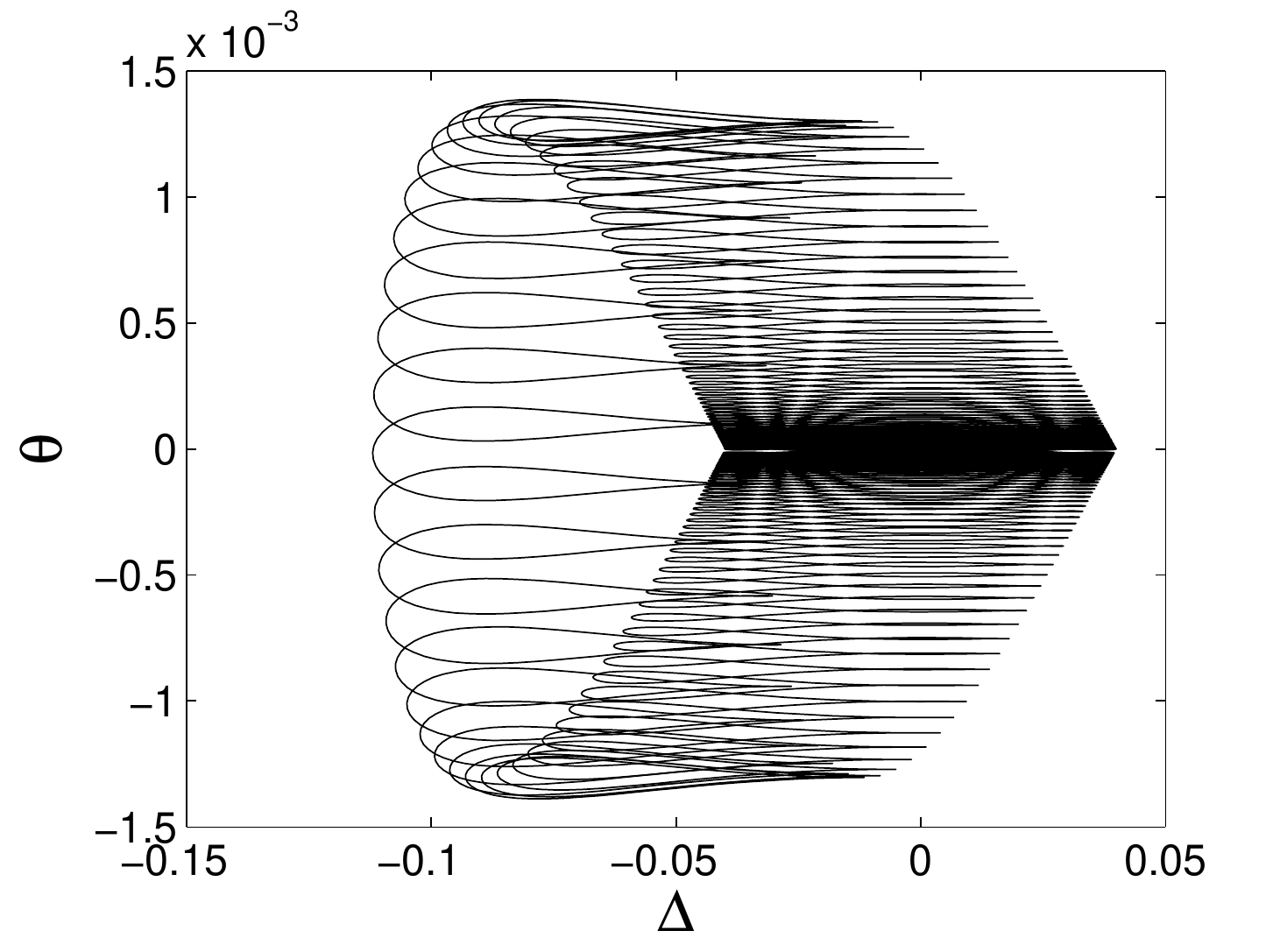}}
\subfigure[]{\includegraphics[width=8cm]{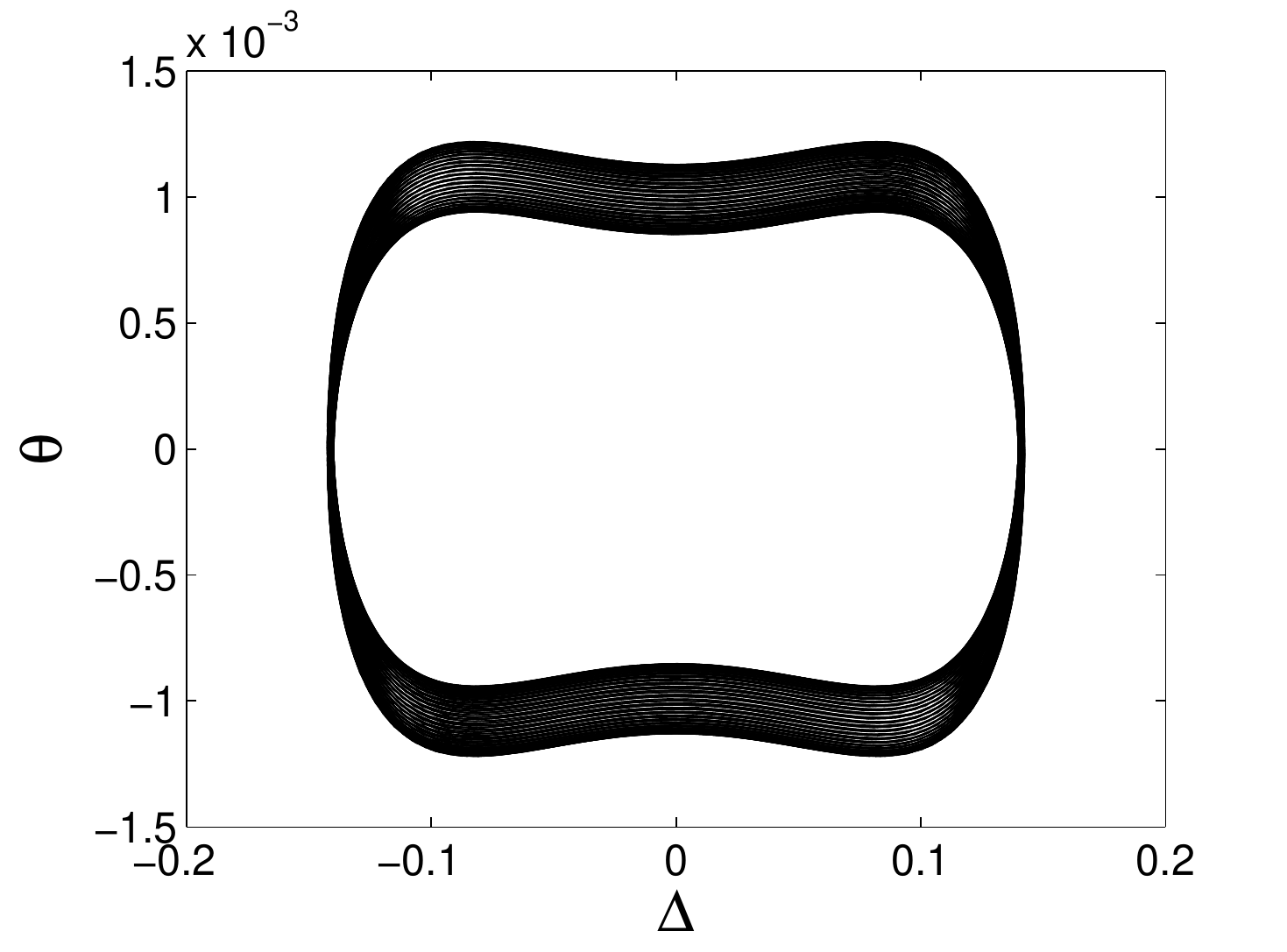}}
\end{center}
\caption{(a) The real and imaginary part of the critical Floquet multipliers shown in black and red solid curves respectively as a function of the solution norm $N$ with $a=0.1$. The dashed curve is our approximation (\ref{ev11}). (b-c) The time-dynamics of the power imbalance $\Delta(z)$ and the phase-difference $\theta(z)$ for the initial condition $\Delta(0)=-Na$, $\theta(0)=0$ and (b) $N=2c=0.4$ and (c) $N=2c=1.4$, corresponding to an unstable and stable symmetric state, respectively.}
\label{fig_7}
\end{figure*}

We consider the equivalent equation \eqref{dth}. Studying the continuation of the symmetric state $(\Delta,\theta)=(0,0)$ in the presence of a parametric drive with small amplitude $|a|\ll1$ at the bifurcation point of the undriven case $c=N/2$, we take the following expansion
\[
\Delta=a\Delta^{(1)}+\mathcal{O}(a^2),\,\theta=a\theta^{(1)}+\mathcal{O}(a^2),
\]
which upon substitution into \eqref{dth} yields
\begin{equation}
\frac{d}{dz}\Delta^{(1)}=-N\theta^{(1)}+N\sin z,\,\frac{d}{dz}\theta^{(1)}=0.
\end{equation}
Looking for periodic solutions, we obtain
\begin{equation}
\Delta^{(1)}=-N\cos z,\,\theta^{(1)}=0.
\end{equation}

Next, we study the stability of the periodic solution above. The linearisation of \eqref{dth} about periodic solution $\Delta$ and $\theta$ is
    \begin{equation}
    \begin{array}{lll}
    \dot{x} &=&-\frac{2\,c\Delta}{\sqrt {1-{\Delta}^{2}}} \left(\sin {x_0} \cos {
\theta} -\cos {x_0} \sin{\theta} \right)x\\
&& -2\,c\sqrt {1-{\Delta}^{2}} \left(\sin {x_0} \sin {
\theta} +\cos {x_0} \cos{\theta} \right)y,
 \\
    \dot{y} &=& \frac{2c}{\sqrt {1-{\Delta}^{2}}}\left(\frac{\Delta^2}{1-\Delta^2}+1\right)\left(\cos {x_0} \cos  {\theta} + \sin {x_0}  \sin {\theta} \right)x\\
    &&-N\delta x
    +\frac {2\Delta c}{\sqrt {1-{\Delta}^{2}}}\left(-\cos {x_0} \sin  {\theta} + \sin {x_0}  \cos {\theta} \right)y.
    \end{array}
    \label{dthev}
    \end{equation}
We will use the solution of the equation to construct a Floquet matrix that will determine whether or not the periodic solution is stable. It is then natural to expand the variables $x(t)$ and $y(t)$ in series as $\Delta$ and $\theta$ above, i.e.\
\begin{equation}
x=x^{(0)}+a^2x^{(2)}+\mathcal{O}(a^3),\, y=y^{(0)}+a^2y^{(2)}+\mathcal{O}(a^3).
\label{series}
\end{equation}
Substituting the expansions into the linearized equation \eqref{dthev}, from terms of $\mathcal{O}(1)$ we obtain
\begin{equation}
\dot{x}^{(0)}=-Ny^{(0)},\,\dot{y}^{(0)}=0.
\label{ev1}
\end{equation}
From terms of $\mathcal{O}(a^2)$ we have
\begin{equation}
\displaystyle
\begin{array}{lll}
\dot{x}^{(2)}&=&-Ny^{(2)}+\frac{N^2}2\sin(2z)x^{(0)}\\
&&+\frac{N}4\left(N^2\left(\cos(2z)+1\right)+1-\cos(2z)\right)y^{(0)},\\
\dot{y}^{(2)}&=&\frac{N}4\left(3N^2-1\right)\left(\cos(2z)+1\right)x^{(0)}-\frac{N^2}2\sin(2z)y^{(0)}.
\end{array}
\label{ev2}
\end{equation}
Equations (\ref{ev1}) are subject to the initial conditions that either $x^{(0)}(0)=1,\,y^{(0)}(0)=0$ or $x^{(0)}(0)=0,\,y^{(0)}(0)=1$, while \eqref{ev2} is solved with the conditions $x^{(2)}(0)=y^{(2)}(0)=0$.

Solving the linear equations for the two sets of initial conditions above and evaluating the values of the functions after one period $T=2\pi$, one will obtain the following Floquet matrix
\begin{widetext}
\begin{eqnarray}
M=\left(
\begin{array}{cc}
1+\frac{a^2}2N^2\pi^2(1-3N^2) & -2N\pi+\frac{N\pi a^2}{6}\left(N^2
\left(4\pi^2+3\right)
+3\right)\\
\frac{a^2}2N\pi(-1+3N^2) & 1+\frac{a^2}2N^2\pi^2(1-3N^2)
\end{array}
\right)+\mathcal{O}(a^3).
\end{eqnarray}
\end{widetext}
Note that the perturbation series \eqref{series} is nonuniform. The approximation above breaks down at $z\sim a^{-3/2}$, which limits the period $T$ for a given drive amplitude $a$.

Calculating the eigenvalues of $M$, we obtain that
\begin{equation}
\lambda_{1,2}=1\pm N\pi a\sqrt{1-3N^2}-\mathcal{O}(a^2).
\label{ev11}
\end{equation}
Because $\lambda>1$ for $0<N<1/\sqrt{3}$, to the leading order we conclude that the continuation of the symmetric state is unstable in the interval and stable otherwise. The stability is in agreement with Fig.\ \ref{fig_3} above. The instability for $N<1/\sqrt{3}$ implies that the symmetry breaking bifurcation occurs earlier than for the undriven case, which we also observed numerically (not shown here). The analytical approximation (\ref{ev11}) is compared to the numerics in Fig.\ \ref{fig_7}(a), where rather perfect agreement is obtained. To illustrate our finding, we show in Fig.\ \ref{fig_7} the time dynamics of the initial condition $\Delta(0)=-Na,\,\theta(0)=0$ for $N=2c=0.4$ and $N=2c=1.4$ with $a=0.1$ corresponding to unstable and stable cases, respectively. The same method can be applied to analyse the existence and stability of the continuation of equilibrium solutions discussed in the previous sections for $|a|\ll1$. However, in general the explicit expression of the solutions as well as the Floquet matrices will be very lengthy.

\section{Conclusion}

Through numerically solving parametrically driven coupled nonlinear Schr\"odinger equations describing the dynamics of wavefields in an oscillating double-well potential, we have shown that such parametric drives may stabilize or destabilize the continuations of equilibrium time-independent states that are respectively unstable or stable in the undriven case. The analysis is performed by employing numerical continuations to find periodic orbits when varying a parameter and by calculating the corresponding Floquet multipliers of the states. Analytical calculations that accompany the numerical results using an averaging method and perturbation expansion have been presented where good quantitative agreement is obtained.

\section*{Acknowledgement}

HJ, HS, and TMB acknowledge the partial financial support of a University of Nottingham Interdisciplinary High Performance Computing (iHPC), School of Mathematics and Faculty of Engineering.

\end{document}